\documentclass[a4paper,11pt]{article}
\pdfoutput=1 % if your are submitting a pdflatex (i.e. if you have
\usepackage{jheppub} % for details on the use of the package, please
\usepackage[style=numeric-comp,
            sorting=none, 
            url = false
]{biblatex}
\usepackage{atlasbiblatex}
\usepackage[T1]{fontenc} % if needed

\usepackage[
    separate-uncertainty = true]
{siunitx}

\usepackage{enumitem}
\usepackage{circledsteps}
\usepackage[font = small]{caption}
\usepackage{subcaption}
\usepackage{multirow}
\usepackage{booktabs}
\usepackage{xspace}
\usepackage{wrapfig}
\usepackage{placeins}
\usepackage{float}
\usepackage[dvipsnames]{xcolor}
\usepackage{ulem}
\usepackage{cancel}

\newcommand{\loose}{\textit{loose}\xspace}
\newcommand{\tight}{\textit{tight}\xspace}
\newcommand{\ntight}{\textit{\mbox{non-tight}}\xspace}
\newcommand{\Nl}{\ensuremath{N_{\mathrm{L}}}\xspace}
\newcommand{\Nt}{\ensuremath{N_{\mathrm{T}}}\xspace}
\newcommand{\Nnt}{\ensuremath{N_{\mathrm{nT}}}\xspace}
\newcommand{\epsreal}{\ensuremath{\varepsilon^{\mathrm{r}}}\xspace}
\newcommand{\epsfake}{\ensuremath{\varepsilon^{\mathrm{f}}}\xspace}

\newcommand{\Ntf}{\ensuremath{N_{\mathrm{T}}^{\mathrm{f}}}\xspace}
\newcommand{\ntf}{\ensuremath{\nu_{\mathrm{T}}^{\mathrm{f}}}\xspace}

\title{Reformulation of a likelihood approach to fake-lepton estimation in the framework of Bayesian inference
}

\author[a]{Johannes Erdmann,}
\author[a]{Cornelius Grunwald,}
\author[a]{Kevin Kröninger,}
\author[a]{Salvatore La Cagnina,}
\author[a]{Lars Röhrig,}
\author[b]{Erich Varnes}
\affiliation[a]{TU Dortmund University, Germany}
\affiliation[b]{The University of Arizona, USA}

\emailAdd{lars.roehrig@tu-dortmund.de}

\abstract{Prompt isolated leptons are essential in many analyses in high-energy particle physics but are subject to fake-lepton background, i.e.~objects that mimic the lepton signature. The fake-lepton background is difficult to estimate from simulation and is often directly determined from data. A popular method is the matrix method, which however suffers from several limitations. This paper recapitulates an alternative approach based on a likelihood with Poisson constraints and reformulates the problem from a different starting point in the framework of Bayesian statistics. The equality of both approaches is shown and several cases are studied in which the matrix method is limited. In addition, the fake lepton background is recalculated and compared to the estimate with the matrix method in an example top-quark measurement.
}

\begin{document}
\maketitle
\flushbottom

% content should go into Content directory
\section{Introduction}
\label{sec:intro}

At hadron colliders, the detector signatures of prompt high-energy electrons and muons can be mimicked by other objects (so-called ``fake leptons"), in particular by particles that are produced in jets. These are for example non-prompt leptons, such as leptons from semileptonic $B$ meson decays, or---for electron fakes---jets with a high electromagnetic fraction and converted photons. The fake-lepton background contribution is difficult to estimate from simulations due to the small misidentification probabilities, meaning that a prohibitively large number of simulated events would need to be generated. It is also difficult to accurately model the fake-lepton contribution in the simulation. Therefore, data-driven techniques are often used instead. Several data-driven methods have been developed, including the matrix method~\cite{PhysRevD.76.092007, Aad:2001975, Aad:1312983}, the ABCD method~\cite{PhysRevD.44.29, Aad:1753190, Aaboud:2299430} and the fake-factor method~\cite{Aaboud:2631950, Aad:2750530, Aaboud:2640802}. In analyses which often search for signal events, the number of estimated background events including contributions from faked leptons is subtracted from the signal event counts. Any bias in the fake-lepton estimation will therefore affect the estimate of signal events, especially if the number of observed events is small. In such cases, it is crucial to have an accurate statistical model to correctly estimate the uncertainty and avoid any bias.

Motivated by limitations of the matrix method, in particular the possibility that the predicted fake-lepton event yields can be negative, a maximum-likelihood approach was proposed in Ref.~\cite{erich_varnes}. In this paper, this improved method is reformulated in Bayesian reasoning, the equality of both approaches is shown and both formulations are implemented in the Bayesian Analysis Toolkit (BAT)~\cite{schulz2020batjl}, a multi-purpose software package for Bayesian inference. 

The paper is structured as follows: In Section~\ref{sec:classical_matrix_method}, the matrix method is introduced in its original form, and its limitations are briefly presented in Section~\ref{sec:error_propagation}. The likelihood ansatz and the Bayesian ansatz are introduced and the equality of the two approaches is shown (Sections~\ref{sec:statistical_model_likelihood}--\ref{sec:Bayesian_model}). The implementation in BAT is described in Section~\ref{sec:implementation}. The paper closes with a discussion of a concrete physics example and a study of several limit cases where the original matrix methods has problems that are addressed with the improved ansatz (Section~\ref{sec:comparison}). Conclusions are presented in Section~\ref{sec:conclusions}.
\section{The matrix method and its limitations}
\label{sec:MM}

The matrix method is a data-driven method to estimate the fake-lepton contribution. While the matrix method works quite well in most use cases, there are some general limitations and restrictions in its application. The main challenges of the matrix method lie in the uncertainty calculation, which is based on a first-order Taylor-series approximation and in the non-vanishing probability to  estimate negative fake-event yields~\cite{Gillam_2014}. The derivations presented below refer to the single-lepton case, although the methods are applicable to multi-lepton final states as well.

\subsection{Matrix method in its original form} \label{sec:classical_matrix_method}

\begin{figure}[h]
    \centering
    \includegraphics[width = 0.25\textwidth]{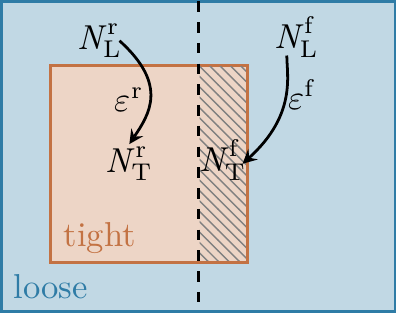}
    \caption{Sketch of the matrix method with the \loose and \tight regions highlighted in blue and red, respectively. The dashed line divides the regions further into real and fake lepton regions.
    \label{fig:MM}}
\end{figure}

In the original formulation of the matrix method~\cite{PhysRevD.76.092007}, two lepton identification criteria are used to estimate the fake-lepton contribution. These criteria are referred to as \loose and \tight. While the \tight requirement corresponds to the region of interest (signal region), i.e.~where the fake-lepton contribution is estimated, the \loose region has relaxed requirements and is enriched in fake leptons. The leptons in the \tight region are a subset of those in the \loose region (Fig.~\ref{fig:MM}).

The corresponding numbers of events passing the \loose and \tight selection are called \Nl and \Nt, containing both, the contribution of real ($N_{\text{L}}^{\text{r}}$ and $N_{\text{T}}^{\text{r}}$) and fake leptons ($N_{\text{L}}^{\text{f}}$ and $N_{\text{T}}^{\text{f}}$). The total event numbers in the regions are then given by
\begin{align}
    \begin{split}
        \Nl &= N_{\text{L}}^{\text{r}} + N_{\text{L}}^{\text{f}}\,, \\
        \Nt &= N_{\text{T}}^{\text{r}} + N_{\text{T}}^{\text{f}}\,.
    \end{split}
    \label{eqn:N_loose_tight}
\end{align}

The probability of migrating from one region to the other is determined by the efficiencies \epsreal and \epsfake for real and fake leptons, respectively\footnote{Differences of these efficiencies caused by lepton flavor, kinematic effects or different sources of fake leptons are neglected in this analysis.}. They are defined as the fraction of leptons passing the corresponding criteria
\begin{align}
    \begin{split}
        \epsreal &= \frac{N_{\text{T}}^{\text{r}}}{N_{\text{L}}^{\text{r}}}\,,\\
        \epsfake &= \frac{N_{\text{T}}^{\text{f}}}{N_{\text{L}}^{\text{f}}}\,.
    \end{split}
    \label{eqn:epsilon_real_fake}
\end{align}
Then, the amount of fake leptons fulfilling the \tight requirements\footnote{It is to note that due to fluctuations in \Nl and \Nt the resulting quantity \Ntf refers to the estimator of the fake-lepton yield in the tight region.} is
\begin{equation}
    \Ntf = \frac{\epsfake}{\epsfake - \epsreal}\cdot\left(\Nt - \epsreal \Nl\right)\,.
    \label{eqn:N_tight_fake}
\end{equation}

This formulation of the matrix method leads to several limitations~\cite{erich_varnes}:
\begin{enumerate}
    \item In the limit of very similar real and fake efficiencies, the estimate becomes numerically unstable, resulting in a large (positive or negative) estimate of \Ntf.
    \item \Ntf becomes negative if the \loose regions contains more events with real leptons than the total number of events in the \tight region.
    \item The uncertainty estimation on the \Ntf yield is subject to limitations that are described in more detail in the next section (Section~\ref{sec:error_propagation}).
\end{enumerate}
In general, the estimation of the fake-lepton background is subject to several sources of systematic uncertainty. While a full discussion of these is beyond the scope of this paper, in principle these uncertainties arise due to potential variations in the values of \epsreal and \epsfake. These variations can be due to differences in the fake-lepton composition between the regions where the \epsfake-values are measured and the analysis region where they are applied, or due to variations in the Monte Carlo model for real-lepton contributions in the fake-lepton control region (since the simulated real-lepton sources are subtracted when measuring \epsfake).

\subsection{The limits of Gaussian uncertainty propagation} \label{sec:error_propagation}

The fake-event yield in the \tight region is subject to uncertainties in the input to Eq.~\eqref{eqn:N_tight_fake}. The uncertainty propagation is based on a first-order Taylor-series approximation (Gaussian uncertainty propagation). Since the efficiencies are given as uncertain parameters to Eq.~\eqref{eqn:N_tight_fake}, the first-order expansion of \Ntf results in
\begin{equation}
    \Ntf \approx \Ntf(\Nl, \Nt, \bar{\varepsilon}^{\mathrm{r}}, \bar{\varepsilon}^{\mathrm{f}}) + \underbrace{\sum_{i \in [\epsreal, \epsfake]} \frac{\partial \Ntf}{\partial x_i} (x_i - \bar{x}_i)}_{\approx\sigma_{\!\Ntf}}\,. 
    \label{eqn:first_order_expansion}
\end{equation}

This corresponds to a linearisation of the output distribution and \Ntf is then assumed to follow a Gaussian distribution. Since Eq.~\eqref{eqn:N_tight_fake} is a non-linear function, a symmetric interval of $\pm\sigma_{\!\Ntf}$ around its mean does not always contain $\SI{68.27}{\percent}$ of the probability-density distribution. This is due to fact that the calculation of uncertainties with Gaussian uncertainty propagation neglects higher-order derivatives and since these do not vanish in Eq.~\eqref{eqn:N_tight_fake}, this effects the accuracy in the estimate of the uncertainty.
\section{An improved matrix method} \label{sec:statistical_model}

In the following, a maximum likelihood approach is described, as motivated and explained in more detail in Ref.~\cite{erich_varnes}. The motivation for this method and the Bayesian derivation is mentioned afterwards. It is shown that both, the likelihood matrix method and the probabilistic derivation are mathematically identical and are further linked with \textit{a-priori} knowledge on the likelihood's parameters in a Bayesian model.

\subsection{Likelihood ansatz} \label{sec:statistical_model_likelihood}

The likelihood matrix method as derived in Ref.~\cite{erich_varnes} describes leptons passing the \loose criteria to be subsequently divided into two orthogonal groups, referred to as \tight and \ntight.

The sum of these two groups is given as the number of events in the \loose sample
\begin{equation}
    \Nl = \Nt + \Nnt\,.
\end{equation}
It follows directly $\Nt\subset\Nl$. The original matrix-method formalism is used to describe the transition into the \tight and \ntight region. Therefore, the entries in both regions are given by
\begin{alignat}{2}
    \Nt &= N_{\mathrm{T}}^{\mathrm{r}} + N_{\mathrm{T}}^{\mathrm{f}}\label{eqn:Nt_lhmm} &&= \epsreal N_{\mathrm{L}}^{\mathrm{r}} + \epsfake N_{\mathrm{L}}^{\mathrm{f}}\,,\\
    \Nnt &= \underbrace{\frac{1 - \epsreal}{\epsreal} N_{\mathrm{T}}^{\mathrm{r}} + \frac{1 - \epsfake}{\epsfake} N_{\mathrm{T}}^{\mathrm{f}}\label{eqn:Nnt_lhmm}}_{\tight\text{\;frame}} &&= \underbrace{(1 - \epsreal) N_{\mathrm{L}}^{\mathrm{r}} + (1 - \epsfake) N_{\mathrm{L}}^{\mathrm{f}}\vphantom{\frac{1 - \epsreal}{\epsreal} N_{\mathrm{T}}^{\mathrm{r}} + \frac{1 - \epsfake}{\epsfake} N_{\mathrm{T}}^{\mathrm{f}}} }_{\loose\text{\;frame}}\,.
\end{alignat}
Since both Eq.~\eqref{eqn:Nt_lhmm} and Eq.~\eqref{eqn:Nnt_lhmm} can be defined in the \tight and the \loose region, there are two different corresponding parameterisations for a maximum-likelihood approach~\cite{erich_varnes}. The efficiencies are parameterised to follow Gaussian distributions $\mathcal{N}$, hence the likelihood results in
\begin{equation}
    p(\Nt, \Nnt, \epsreal, \epsfake | \nu_{\mathrm{T}}, \nu_{\mathrm{nT}}, \hat{\varepsilon}^{\mathrm{r}}, \hat{\varepsilon}^{\mathrm{f}}) = p(\Nt | \nu_{\mathrm{T}}) \cdot p(\Nnt | \nu_{\mathrm{nT}}) \cdot \mathcal{N}(\epsreal, \sigma_{\varepsilon^{\mathrm{r}}} | \hat{\varepsilon}^{\mathrm{r}}) \cdot \mathcal{N}(\epsfake, \sigma_{\varepsilon^{\mathrm{f}}} | \hat{\varepsilon}^{\mathrm{f}})\,,
    \label{eqn:lhmm_likelihood_1}
\end{equation}
with Poissonian constraints on \Nt and \Nnt and corresponding estimators $\nu_{\mathrm{T}}$ and $\nu_{\mathrm{nT}}$. In order to simplify the fit, the uncertainties on the efficiencies are assumed to be negligible compared to the Poissonian uncertainties \footnote{It is possible to re-run the fit with variations in the efficiency to estimate a systematic uncertainty for the estimate.}, so the efficiencies are parameterised as $\delta$ distributions in Ref.~\cite{erich_varnes}. The likelihood from Eq.~\eqref{eqn:lhmm_likelihood_1} is then simplified to
\begin{equation}
    p(\Nt, \Nnt, \epsreal, \epsfake | \nu_{\mathrm{T}}, \nu_{\mathrm{nT}}, \hat{\varepsilon}^{\mathrm{r}}, \hat{\varepsilon}^{\mathrm{f}}) = p(\Nt | \nu_{\mathrm{T}}) \cdot p(\Nnt | \nu_{\mathrm{nT}})\,.
    \label{eqn:lhmm_likelihood_2}
\end{equation}

In the following, the motivation of the likelihood matrix method is presented and the equality of the likelihood and Bayesian approaches is shown. The ansatz is made up of two parts, each returning a probability that the lepton is real or fake in the \loose and \tight regions. Beginning with the \loose contribution, the parameters of the likelihood that are used to express the estimators of \loose real and fake leptons are referred to as $\nu^{\mathrm{r}}$ and $\nu^{\mathrm{f}}$.

As the event yields are assumed to follow a Poisson distribution, the likelihood for the real and fake-event yields is given by
\begin{equation}
    p(N_{\mathrm{L}}^{\mathrm{r/f}} | \nu^{\mathrm{r/f}}) = \mathrm{Poisson}(N_{\mathrm{L}}^{\mathrm{r/f}} | \nu^{\mathrm{r/f}})\,.
\end{equation}
Since neither $N_{\mathrm{L}}^{\mathrm{r}}$ nor $N_{\mathrm{L}}^{\mathrm{f}}$ are measured, but only their sum \Nl, the likelihood for the \loose region must contain a sum over the possible splits of \Nl into $N_{\mathrm{L}}^{\mathrm{r}}$ and $N_{\mathrm{L}}^{\mathrm{f}}$:

\begin{equation}
    p(\Nl | \nu^{\mathrm{r}}, \nu^{\mathrm{f}}) = \sum_{x = 0}^{\Nl} p(\Nl - x | \nu^{\mathrm{r}}) \cdot p(x | \nu^{\mathrm{f}})\,.
\end{equation}
The sum index $x$ counts up all $N_{\mathrm{L}}^{\mathrm{f}}$ until \Nl is reached.

The migration from \loose to \tight is defined as the number of successes when the lepton passes or fails the \tight requirements. Therefore, a binomial distribution is used as the probability density function. The probability is given by the efficiencies \epsreal and \epsfake for real and fake leptons, respectively. The corresponding estimators for the efficiencies are indicated as $\hat{\varepsilon}^i$. The process is described via
\begin{equation}
    p(N_{\mathrm{T}}^{\mathrm{r/f}} | \hat{\varepsilon}^{\mathrm{r/f}}, N_{\mathrm{L}}^{\mathrm{r/f}}) = \mathrm{Binomial}(N_{\mathrm{T}}^{\mathrm{r/f}} | \hat{\varepsilon}^{\mathrm{r/f}}, N_{\mathrm{L}}^{\mathrm{r/f}})\,.
\end{equation}

The given number of \loose real and fake leptons is provided to the binomial distribution as output of the Poisson distribution. Again, only the overall number of \tight leptons can be measured and hence a second sum must be used to provide the total probability for a set of data $\mathcal{D} = (\Nl, \Nt, \epsreal, \epsfake)$ given the parameter vector $\boldsymbol{\theta} = (\nu^{\mathrm{r}}, \nu^{\mathrm{f}}, \hat{\varepsilon}^{\mathrm{r}}, \hat{\varepsilon}^{\mathrm{f}})$
\begin{equation}
    p(\mathcal{D} | \boldsymbol{\theta}) = \sum_{x = 0}^{\Nl} \sum_{y = y_{\mathrm{min}}}^{y_{\mathrm{max}}} p(\Nl - x | \nu^{\mathrm{r}}) \cdot p(x | \nu^{\mathrm{f}}) \cdot p(\Nt - y | \hat{\varepsilon}^{\mathrm{r}}, \Nl - x) \cdot p(y | \hat{\varepsilon}^{\mathrm{f}}, x)\,.
    \label{eqn:likelihood_first_principles}
\end{equation}
Since the binomial terms $p(\Nt - y | \hat{\varepsilon}^{\mathrm{r}}, \Nl - x)$ and $p(y | \hat{\varepsilon}^{\mathrm{f}}, x)$ describe the migration from the \loose to the \tight region, it is only effected by the measured values given the probability estimators $\hat{\varepsilon}^{\mathrm{r/f}}$.
The sum index $y$ counts up all \Ntf. The upper limit $y_{\mathrm{max}}$ is due to limitations that $N_{\mathrm{T}}^{\mathrm{r/f}}$ cannot be greater than either \Nt nor $N_{\mathrm{L}}^{\mathrm{r/f}}$, i.e.~the range is given as
\begin{equation}
    N_{\mathrm{T}}^{\mathrm{r/f}} \leq \mathrm{min}(\Nt, N_{\mathrm{L}}^{\mathrm{r/f}}) = y_{\mathrm{max}}\,.
    \label{eqn:sum_index_1}
\end{equation}

Since $N_{\mathrm{L/T}}^{\mathrm{r}} = N_{\mathrm{L/T}} - N_{\mathrm{L/T}}^{\mathrm{f}}$, Eq.~\eqref{eqn:sum_index_1} can be expressed as
\begin{equation}
    N_{\mathrm{T}}^{\mathrm{r}} \leq \mathrm{min}(\Nt, N_{\mathrm{L}}^{\mathrm{r}}) = \mathrm{min}(\Nt, \Nl - N_{\mathrm{L}}^{\mathrm{f}})\,.
    \label{eqn:sum_index_2}
\end{equation}
From Eq.~\eqref{eqn:sum_index_2}, \Ntf has to be greater or equal to $\Nt - \mathrm{min}(\Nt, \Nl - N_{\mathrm{L}}^{\mathrm{f}})$. This relation is true if
\begin{equation}
    \Ntf \geq \mathrm{max}(0, \Nt - \Nl + \underbrace{N_{\mathrm{L}}^{\mathrm{f}}}_{= x}) = y_{\mathrm{min}}\,.
\end{equation}
Since $y$ counts up all \tight and fake leptons, the sum indices $y_{\mathrm{min}}$ and $y_{\mathrm{max}}$ are defined as
\begin{align}
    \begin{split}
        y_{\mathrm{min}} &= \mathrm{max}(0, \Nt - \Nl + x)\,,\\
        y_{\mathrm{max}} &= \mathrm{min}(\Nt, x)\,.
    \end{split}
\end{align}
Though Eq.~\eqref{eqn:lhmm_likelihood_2} and Eq.~\eqref{eqn:likelihood_first_principles} were derived with different approaches, they are in fact equivalent, as proven in Appendix~\ref{sec:appendix_proof}.

\subsection{Bayesian model} \label{sec:Bayesian_model}

Prior knowledge is an essential part of Bayesian inference. Physical knowledge flows into the choice of the prior in order to obtain an appropriate posterior distribution. 
Thus this choice leads to a certain degree of subjectivity.

It was shown that both likelihoods are mathematically identical, therefore, the formalism of the likelihood matrix method (see Eq.~\eqref{eqn:lhmm_likelihood_2}) is used due to smaller computation time compared to the likelihood with the two sums in Eq.~\eqref{eqn:likelihood_first_principles}. 
As mentioned before, the efficiencies are chosen to follow $\delta$ distributions in Ref.~\cite{erich_varnes}. Since the formulation of the model in Eq.~\eqref{eqn:likelihood_first_principles} considers the uncertainties as free parameters of the model, they are constrained by the prior below. 

The prior distributions in the case of fake-lepton estimation are chosen to be as uninformative as possible, but at the same time physical. Here, $\nu^{\mathrm{r}}$ and $\nu^{\mathrm{f}}$ follow uniform distributions with ranges from $[0, n_{\nu}\cdot\Nl]$ due to a small but non-vanishing probability for $\nu^{\mathrm{r/f}}$ to be larger than \Nl. The concrete value of the scale factor $n_{\nu}$ for the prior range can be chosen specifically for each use case to account for the full posterior probability density of the parameters $\nu^{\mathrm{r/f}}$.

The efficiencies are assumed to follow a truncated Normal distribution~\cite{Robert_1995} $\mathcal{N}_{\!\text{tr}}$ from $a = 0$ to $b = 1$ with mean $\varepsilon$ and standard deviation $\sigma_{\varepsilon}$, defined by
\begin{equation}
  \mathcal{N}_{\!\text{tr}}(\varepsilon, \sigma_{\varepsilon}) = \frac{\phi\left( \frac{x - \varepsilon}{\sigma_{\varepsilon}} \right)}{\mathrm{\Phi}\left(\frac{b - \varepsilon}{\sigma_{\varepsilon}}\right) - \mathrm{\Phi}\left(\frac{a - \varepsilon}{\sigma_{\varepsilon}}\right)}\,,
  \label{eqn:truncated_normal}
\end{equation}
where the probability density function $\phi(\alpha)$ and the cumulative distribution function $\mathrm{\Phi}(\beta)$ are defined by
\begin{align}
\begin{split}
  \phi(\alpha) &= \frac{1}{\sqrt{2\pi\sigma^2}}\cdot\mathrm{e}^{-\frac{\alpha^2}{2}}\,,\\
  \mathrm{\Phi}(\beta) &= \frac{1}{2}\cdot\left(1 +  \mathrm{erf}\left(\frac{\beta}{\sqrt{2}}\right)\right)\,.
\end{split}
\end{align}
Other choices for the prior distributions can of course be made.

To develop the likelihood matrix method as a Bayesian model, the likelihood is parameterised according to Eq.~\eqref{eqn:Nt_lhmm} and Eq.~\eqref{eqn:Nnt_lhmm} and the priors for the efficiencies follow the truncated Normal distributions. This results in two possible parameterisations
\begin{align}
    \begin{split}
        \boldsymbol{\theta}_{\mathrm{L}}^{\mathrm{LHMM}} &= (\nu^{\mathrm{r}}, \nu^{\mathrm{f}}, \hat{\varepsilon}^{\mathrm{r}}, \hat{\varepsilon}^{\mathrm{f}}) = \boldsymbol{\theta}\,, \\
        \boldsymbol{\theta}_{\mathrm{T}}^{\mathrm{LHMM}} &= (\nu_{\mathrm{T}}^{\mathrm{r}}, \nu_{\mathrm{T}}^{\mathrm{f}}, \hat{\varepsilon}^{\mathrm{r}}, \hat{\varepsilon}^{\mathrm{f}})\,, \\
    \end{split}
\end{align}
which refer to the parameters in the likelihood in Eq.~\eqref{eqn:lhmm_likelihood_2}.
With the additional consideration of uncertainties in the efficiencies, the final likelihood is updated to
\begin{equation}
    p(\Nt, \Nnt, \epsreal, \epsfake | \nu_{\mathrm{T}}, \nu_{\mathrm{nT}}, \hat{\varepsilon}^{\mathrm{r}}, \hat{\varepsilon}^{\mathrm{f}}) = p(\Nt | \nu_{\mathrm{T}}) p(\Nnt | \nu_{\mathrm{nT}})\mathcal{N}_{\!\text{tr}}(\epsreal, \sigma_{\epsreal} | \hat{\varepsilon}^{\mathrm{r}})\mathcal{N}_{\!\text{tr}}(\epsfake, \sigma_{\epsfake} | \hat{\varepsilon}^{\mathrm{f}})\,.
    \label{eqn:lhmm_likelihood_3}
\end{equation}
The count rate estimators $\nu$ are defined according to Eq.~\eqref{eqn:Nt_lhmm} and Eq.~\eqref{eqn:Nnt_lhmm} as
\begin{alignat}{2}
    \nu_{\mathrm{T}} &= \nu_{\mathrm{T}}^{\mathrm{r}} + \nu_{\mathrm{T}}^{\mathrm{f}} &&= \hat{\varepsilon}^{\mathrm{r}} \nu^{\mathrm{r}} + \hat{\varepsilon}^{\mathrm{f}} \nu^{\mathrm{f}}\,,\\
    \nu_{\mathrm{nT}} &= \underbrace{\frac{1 - \hat{\varepsilon}^{\mathrm{r}}}{\hat{\varepsilon}^{\mathrm{r}}} \nu_{\mathrm{T}}^{\mathrm{r}} + \frac{1 - \hat{\varepsilon}^{\mathrm{f}}}{\hat{\varepsilon}^{\mathrm{f}}} \nu_{\mathrm{T}}^{\mathrm{f}}}_{\boldsymbol{\theta}_{\mathrm{T}}^{\mathrm{LHMM}} \text{\,parameterisation}} &&= \underbrace{(1 - \hat{\varepsilon}^{\mathrm{r}})\nu^{\mathrm{r}} + (1 - \hat{\varepsilon}^{\mathrm{f}})\nu^{\mathrm{f}} \vphantom{\frac{1 - \hat{\varepsilon}^{\mathrm{r}}}{\hat{\varepsilon}^{\mathrm{r}}} \nu_{\mathrm{T}}^{\mathrm{r}} + \frac{1 - \hat{\varepsilon}^{\mathrm{f}}}{\hat{\varepsilon}^{\mathrm{f}}} \nu_{\mathrm{T}}^{\mathrm{f}}} }_{\boldsymbol{\theta}_{\mathrm{L}}^{\mathrm{LHMM}}\text{\,parameterisation}}\,.
\end{alignat}
Since the \loose parameterisation, indicated with the subscript $_{\mathrm{L}}$, is identical to the parameterisation of the likelihood in Eq.~\eqref{eqn:likelihood_first_principles} and their equality was shown before, the \loose parameterisation is chosen for further studies.
\par\bigskip
Bayesian inference now returns probabilistic statements on the parameters $\boldsymbol{\theta}$ by considering the data $\mathcal{D}$. These statements result from Bayes' theorem~\cite{Bayes_theorem_original}
\begin{equation}
    p(\boldsymbol{\theta}|\mathcal{D}) = \frac{\overbrace{p(\mathcal{D}|\boldsymbol{\theta})}^{\text{Likelihood}} \overbrace{p(\boldsymbol{\theta})}^{\text{Prior}}}{\underbrace{\int p(\mathcal{D}|\boldsymbol{\theta}) p(\boldsymbol{\theta})\,\text{d}\boldsymbol{\theta}}_{\text{Evidence}}}\,.
\end{equation}

Knowledge and inference about certain parameter distributions are determined using the multidimensional posterior probability distribution~\cite{schulz2020batjl} and the integral over all nuisance parameters
\begin{equation}
    p(\theta_i | \mathcal{D}) = \int p(\boldsymbol{\theta} | \mathcal{D}) \prod_{i\neq j}\,\text{d}\theta_i\,.
\end{equation}
The posterior probability for the \tight, fake yield results from the multiplication of both marginalised distributions $p(\nu^{\mathrm{f}} | \mathcal{D})$ and $p(\hat{\varepsilon}^{\mathrm{f}} | \mathcal{D})$
\begin{equation}
    p(\nu_{\mathrm{T}}^{\mathrm{f}} | \mathcal{D}) = p(\nu^{\mathrm{f}} | \mathcal{D}) \cdot p(\hat{\varepsilon}^{\mathrm{f}} | \mathcal{D})\,.
    \label{eqn:nu_tight_fake}
\end{equation}

\subsection{Implementation in the Bayesian Analysis Toolkit}
\label{sec:implementation}

The posterior probability density is built in the \texttt{julia}~\cite{Julia-2017} package \texttt{BAT.jl}~\cite{schulz2020batjl}. It allows for the implementation of statistical models in a Bayesian framework and the inference of their free parameters. It provides a toolkit with numerical algorithms for sampling, optimisation and integration.

BAT.jl currently offers a choice of two main Markov Chain Monte Carlo (MCMC) algorithms in addition to three importance samplers. MCMC algorithms (Metropolis-Hastings~\cite{Metropolis:1953am, Hastings} and Hamiltonian Monte Carlo~\cite{Hamiltonian_MC}) are well suited for high-dimensional parameter-space sampling, while the importance samplers are an easy and fast-to-use alternative in low-dimensional parameter spaces. They are called via the \texttt{bat\_sample} function and provide adjustable, algorithm-specific arguments. Applied for the posterior density to estimate fake-event yields in the \tight region, the Metropolis-Hastings algorithm is the algorithm of choice for $\Nl > 50$, since the importance samplers provide too few samples to fully explore the parameter space in these regions with sufficient resolution.
\section{A comparison of the original and improved matrix methods}\label{sec:comparison}

As an example for the comparison of the original and improved matrix method, the estimation of the fake-lepton contributions in  an inclusive cross-section measurement in top-quark pair production in association with a photon ($t\bar{t}\gamma$) with the ATLAS experiment~\cite{Aaboud_2019} is discussed in Section~\ref{sec:physics_example}. This measurement used data taken in proton--proton collisions at the Large Hadron Collider (LHC) in the years 2015 and 2016, corresponding to an integrated luminosity of \SI{36}{\per\femto\barn}. The matrix method was used to estimate the fake-lepton yield in the single-lepton channel, which mainly stems from multijet background processes. 

Section~\ref{sec:limit_cases} is devoted to the limitations of the matrix method in certain regions of the phase space, as explained in Section~\ref{sec:classical_matrix_method}. Therefore, five situations are discussed and the estimates are compared between the original and improved matrix methods.

The following studies are performed using the likelihood from Eq.~\eqref{eqn:lhmm_likelihood_2} and the Metropolis Hastings algorithm with a sample size of \num{e6}.

\subsection[Physics example: \texorpdfstring{$t\bar{t}\gamma$}{} cross-section measurement with the ATLAS detector]{Physics example: $\boldsymbol{t\bar{t}\gamma}$ cross-section measurement with the ATLAS detector}\label{sec:physics_example}

The efficiencies \epsreal and \epsfake were determined with the tag-and-probe technique, using leptons from $Z$ boson decays and control regions enriched with fake leptons for the real and fake efficiency, respectively~\cite{Aaboud_2019}. 

Since Ref.~\cite{Aaboud_2019} only provides the number of estimated fake events as well as its uncertainty to be $\Ntf = \num{360(200)}$ and \Nt to be $\Nt = \num{11750}$, but does not give \Nl or the real and fake efficiencies, they are estimated here in the following way:

The efficiencies are roughly estimated from Ref.~\cite{ATLAS-CONF-2014-058}. Although they are a function of various kinematic quantities and the uncertainties in each bin include a combination of systematic and statistical uncertainties, the efficiencies are taken to be $\epsreal = \num{0.8}$ and $\epsfake = \num{0.2}$. To reproduce the uncertainty on the fake-event yield $\sigma_{\Ntf} = \num{200}$, Eq.~\eqref{eqn:first_order_expansion} is used to estimate symmetric uncertainties on the efficiencies. For simplicity, the uncertainties on 
\begin{figure}[H]
    \centering
    \includegraphics[width = 0.97\textwidth]{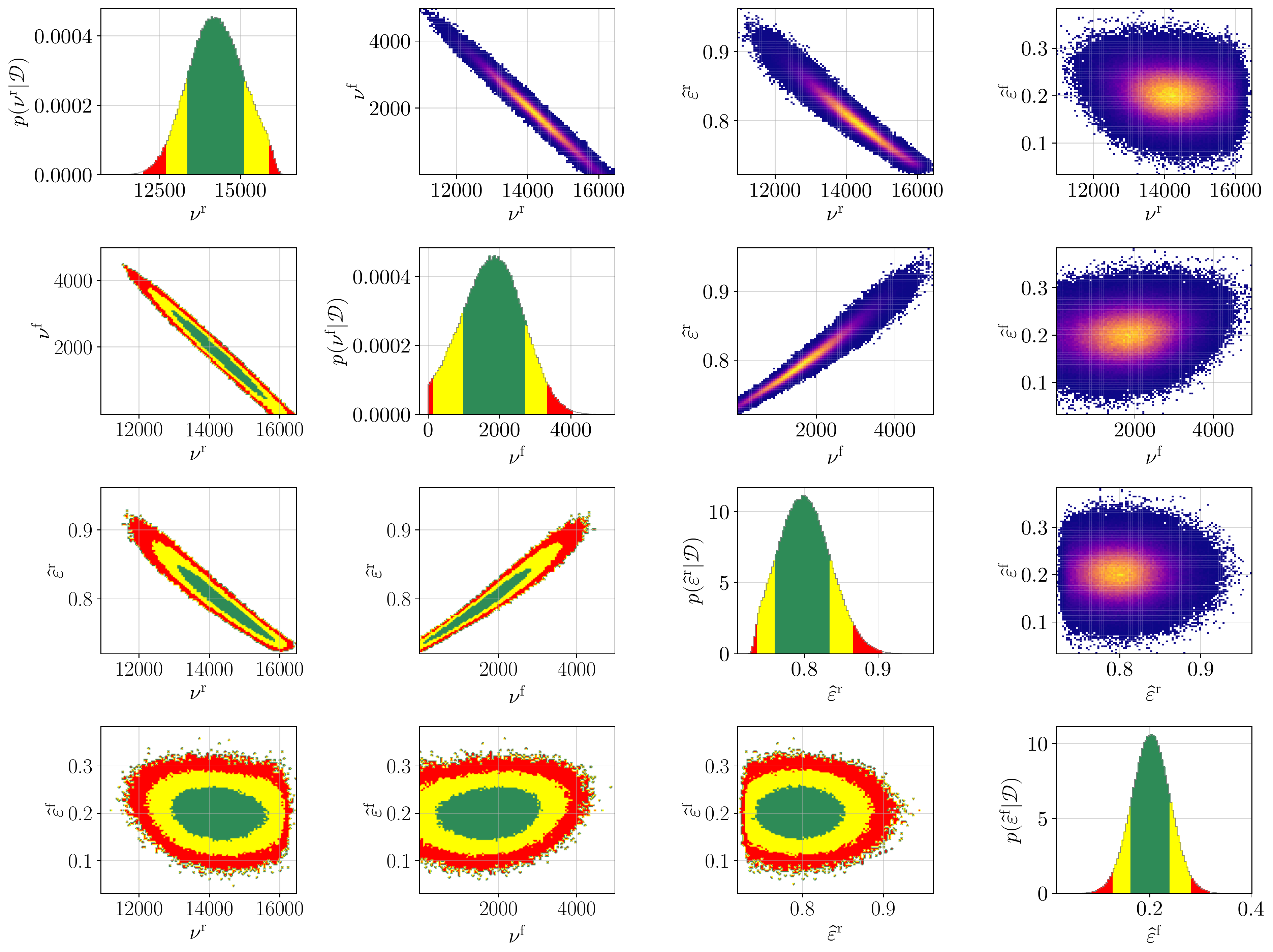}
    \caption{The marginalised distributions of the parameter set $\boldsymbol{\theta}$. The coloured regions belong to the smallest \SI{68.27}{\percent}, \SI{95.45}{\percent} and \SI{99.73}{\percent} intervals. The off-diagonals present two-dimensional histograms of the parameter combinations. They are mirrored on the diagonal with the upper right plots showing heat maps and the lower left plots sharing the plot style and interval definitions of the diagonal histograms.}
    \label{fig:physics_example_md}
\end{figure}
the efficiencies are assumed to be equal, so that $\sigma_{\epsreal} = \sigma_{\epsfake} = \sigma_{\varepsilon}$. Hence, it follows
\begin{equation}
    \sigma_{\varepsilon} = \alpha\sigma_{\Ntf}\,,
\end{equation}
with a factor $\alpha$. The factor is calculated to $\alpha \approx 0.00019$ using an arbitrary $\sigma_{\varepsilon}$ and the resulting uncertainty $\sigma_{\Ntf}$ computed with Eq.~\eqref{eqn:first_order_expansion}. Therefore, the uncertainty on the efficiencies resulting in the output uncertainty of the original matrix method $\sigma_{\Ntf} = 200$ is computed to $\sigma_{\varepsilon} \approx 0.038$.

The missing number of \Nl is derived by transforming Eq.~\eqref{eqn:N_tight_fake} as
\begin{equation}
    \Nl = \frac{1}{\epsreal} \left(\Nt - \frac{\Ntf (\epsfake - \epsreal)}{\epsfake}\right) = \num{16038} \,.
\end{equation}

\begin{figure}[H]
    \centering
    \includegraphics[width = 0.8\textwidth]{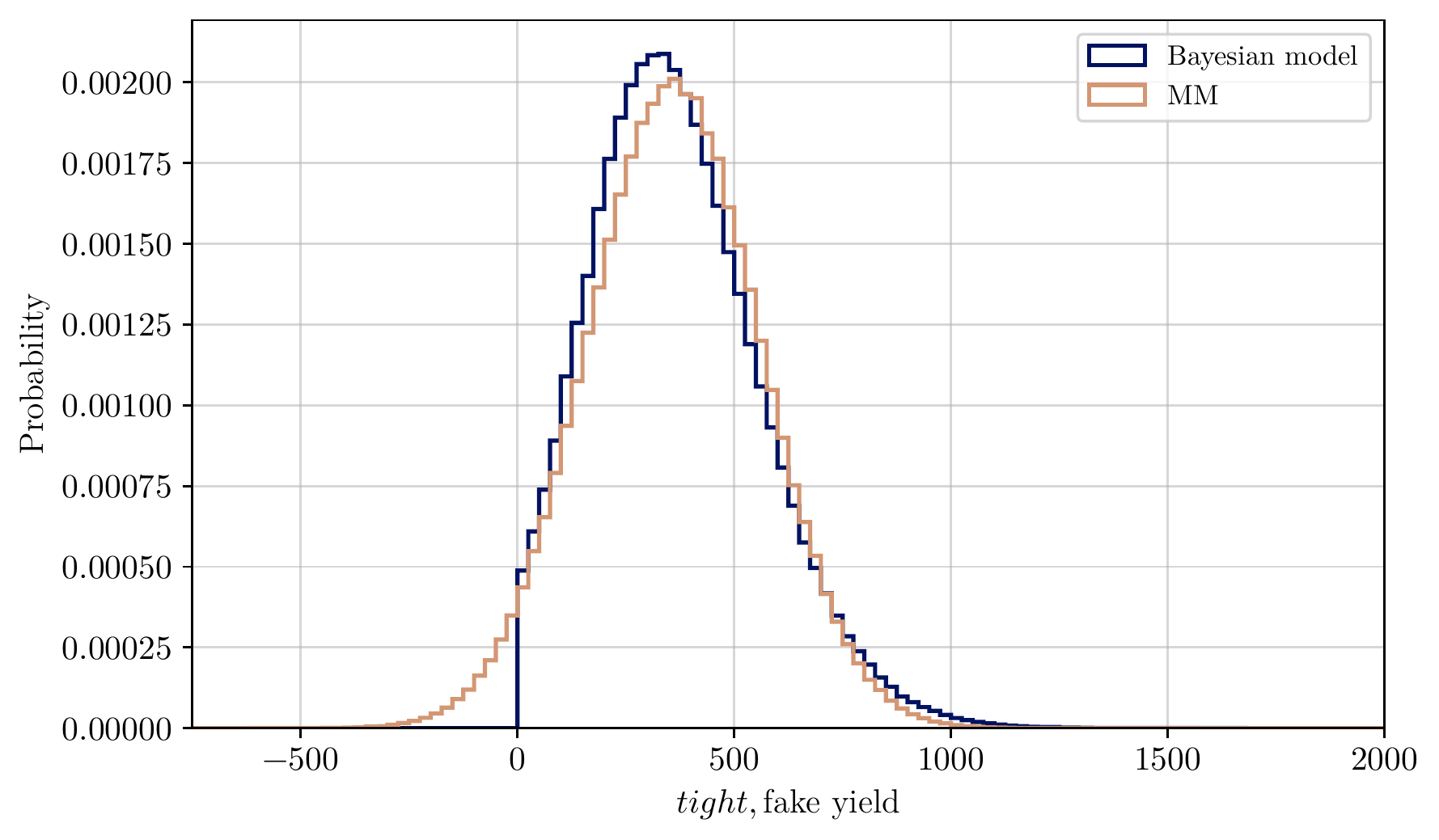}
    \caption{Probability distribution function for the fake event estimator passing the \tight selection in the example of the $t\bar{t}\gamma$ cross-section measurement.}
    \label{fig:physics_example}
\end{figure}

The marginalised distributions of the Bayesian model for the parameter set $\boldsymbol{\theta}$ are shown in Figure~\ref{fig:physics_example_md}. On the main diagonal the posterior parameter distributions are shown with the according \SI{68.27}{\percent}, \SI{95.45}{\percent} and \SI{99.73}{\percent} smallest intervals. The off-diagonals show two-dimensional histograms of the parameter configurations. 
The posterior distribution for fake-lepton estimation results from Eq.~\eqref{eqn:nu_tight_fake} and is shown in Figure~\ref{fig:physics_example} as dark blue histogram. The smallest intervals are omitted here.

In Figure~\ref{fig:physics_example}, the distribution of the original matrix method is shown as well. The estimate with the original matrix method in the orange histogram leads to a non-negligible probability in the unphysical region where the fake yield is negative, while the Bayesian approach does not. Also the values of maximal probability (mean for the original matrix method and mode for the Bayesian approach) differ. The results are summarised in Table~\ref{tab:physics_example}.

\begin{table}[H]
    \small
    \centering
    \caption{Summary of mean values of the distributions in Figure~\ref{fig:physics_example}. The uncertainties refer to the smallest \SI{68.27}{\percent} intervals, since the distribution of the Bayesian model has an asymmetric shape.}
    \begin{tabular}{ccc}
        \toprule
        Quantity & Bayesian model & Original matrix method \\
        \midrule
        mean & \hrulefill & $\num{360\pm200}$ \\
        mode & $300^{+225}_{-175}$ & $\num{360\pm200}$ \\
        median & $352$ & $360$ \\
        \bottomrule
    \end{tabular}
    \label{tab:physics_example}
\end{table}

\subsection{Special cases of the original matrix method}\label{sec:limit_cases}

One of the main problems of the original formulation of the matrix method are the negative event-yield estimates in some regions of the phase space. Thus, five artificial scenarios only based on the measured quantities \Nl, \Nt, \epsreal and \epsfake emphasising the limitations of the matrix method are presented below. Unless otherwise stated, $\Nl = 20$ and $\Nt = 10$.

\begin{description}[leftmargin = 1.6em, font=\normalfont]
    \item[\Circled{1}] The first case targets the numerical instabilities of the matrix method when sampling in a region of the phase space with similar efficiencies. For this example, the efficiencies are chosen to be $\epsreal = \num{0.51\pm0.02}$ and $\epsfake = \num{0.50\pm0.02}$.
    \item[\Circled{2}] The second case demonstrates how the estimation performs when the parameters of the matrix method are chosen so that $\Ntf \approx 0$. This is achieved for a very small fake efficiency or a very small $N_{\mathrm{L}}^{\mathrm{f}}$. For this example the fake efficiency takes on a value close to zero, so that $\epsreal = \num{0.75\pm0.02}$ and $\epsfake = \num{0.01\pm0.02}$.
    \item[\Circled{3}] Since analytically there is a possibility that the matrix method estimates negative fake-event yields, the parameters of Eq.~\eqref{eqn:N_tight_fake} are chosen so that $\Ntf < 0$.
    \item[\Circled{4}] This scenario discusses two aspects. With the efficiencies are chosen to be $\epsreal = \num{0.99\pm0.02}$ and $\epsfake = \num{0.01\pm0.02}$, the very large difference $\mathrm{\Delta}\varepsilon = |\epsreal - \epsfake|$ is expected to lead to a stable estimate with the matrix method. On the other hand, the Bayesian model is evaluated at the prior bounds, as these are bounded below and above with zero and one (see Eq.~\eqref{eqn:truncated_normal}).
    \item[\Circled{5}] The last case examines the effect of the efficiency uncertainty, hence $\sigma_{\varepsilon}$ is increased by a factor of \num{10} and the efficiencies are chosen to be $\epsreal = \num{0.75\pm0.2}$ and $\epsfake = \num{0.42\pm0.2}$.
\end{description}

\begin{figure}[ht]
    \centering
    \begin{subfigure}{0.48\textwidth}
        \centering
        \includegraphics[width = 1.0\textwidth]{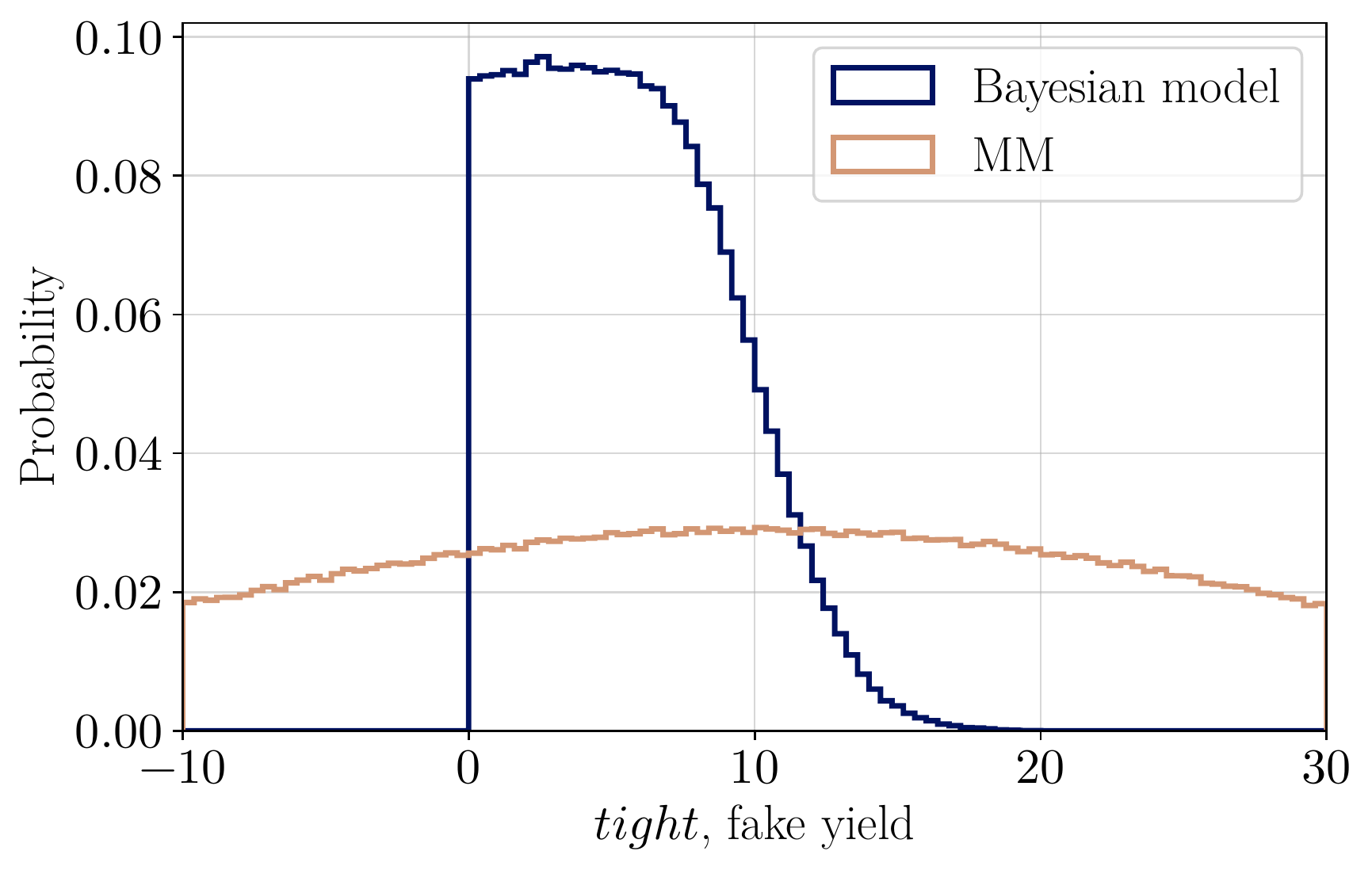}
        \subcaption{Case \Circled{1}.}
        \label{fig:case_1}
    \end{subfigure}
    \begin{subfigure}{0.48\textwidth}
        \centering
        \includegraphics[width = 1.0\textwidth]{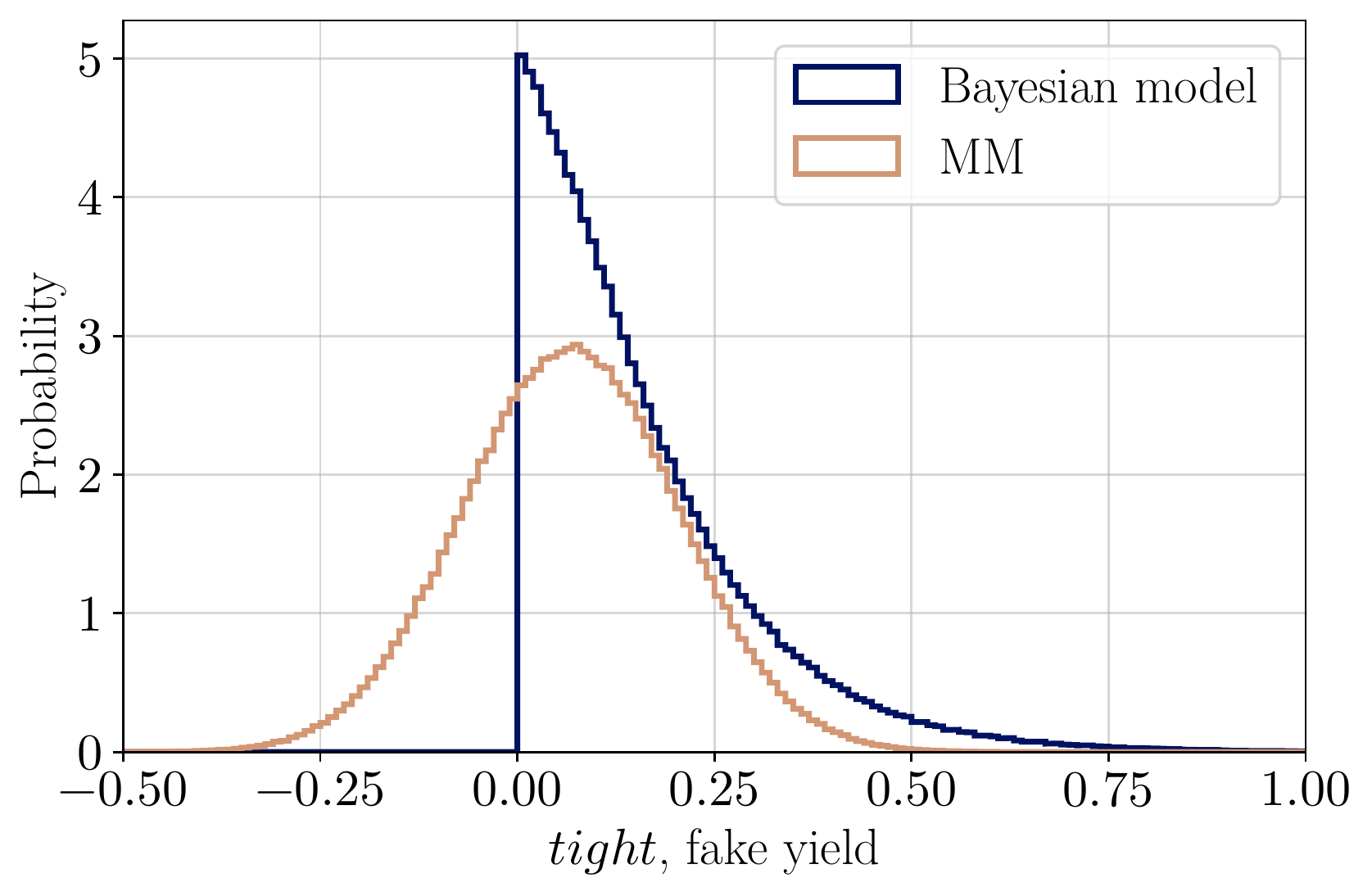}
        \subcaption{Case \Circled{2}.}
        \label{fig:case_2}
    \end{subfigure}\\
    \begin{subfigure}{0.48\textwidth}
        \centering
        \includegraphics[width = 1.0\textwidth]{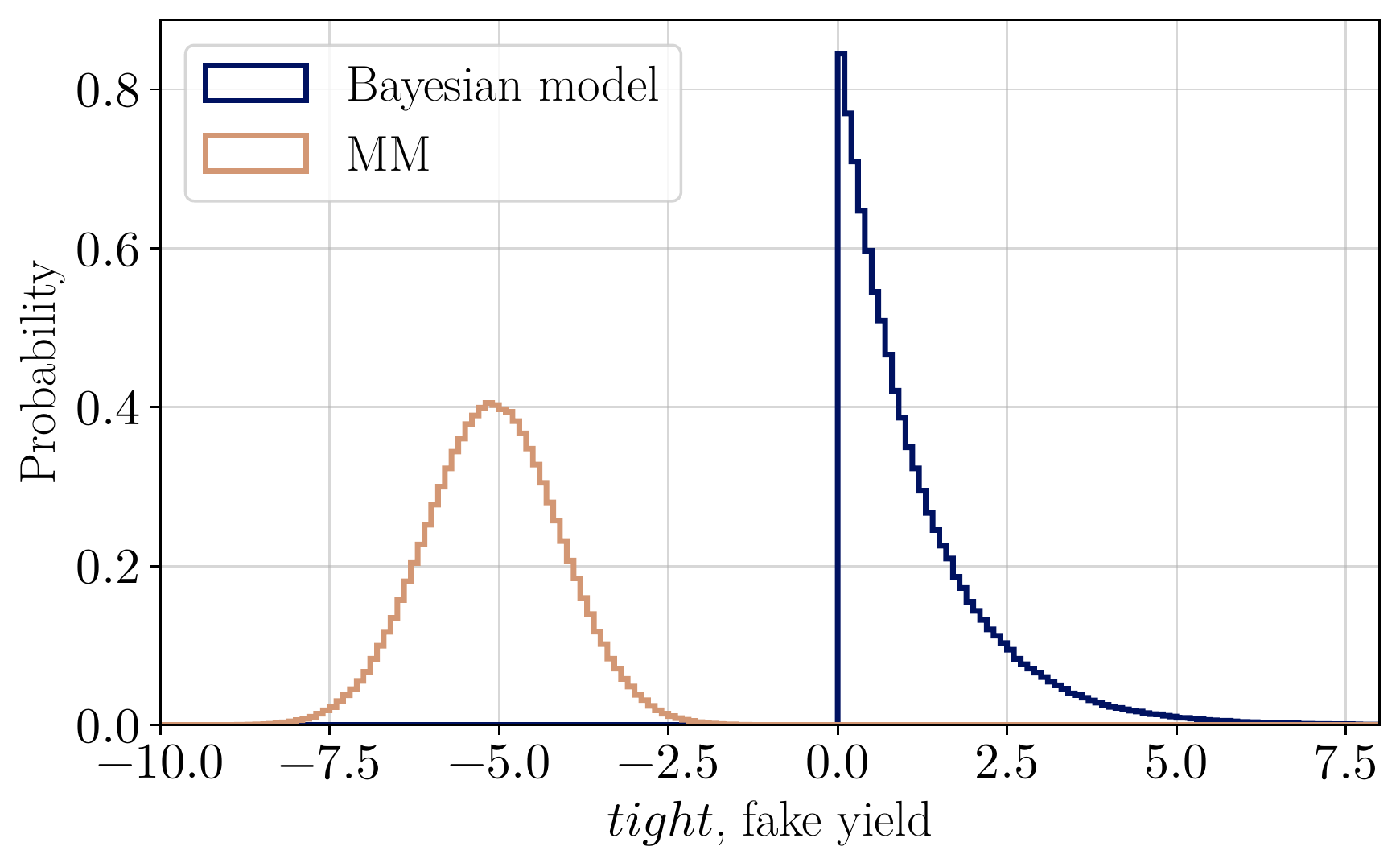}
        \subcaption{Case \Circled{3}.}
        \label{fig:case_3}
    \end{subfigure}
    \begin{subfigure}{0.48\textwidth}
        \centering
        \includegraphics[width = 1.0\textwidth]{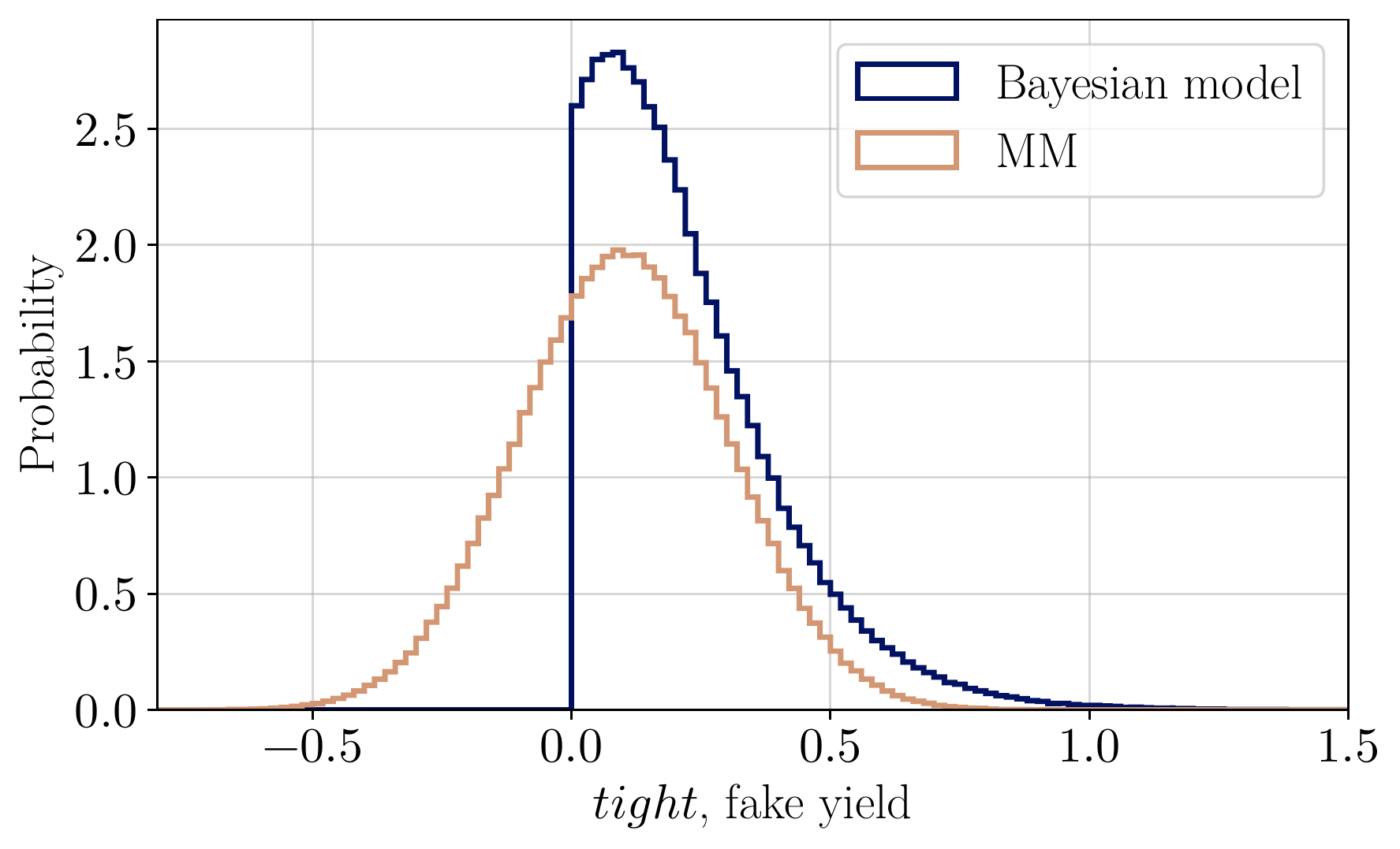}
        \subcaption{Case \Circled{4}.}
        \label{fig:case_4}
    \end{subfigure}\\
    \begin{subfigure}{1\textwidth}
        \centering
        \includegraphics[width = 0.48\textwidth]{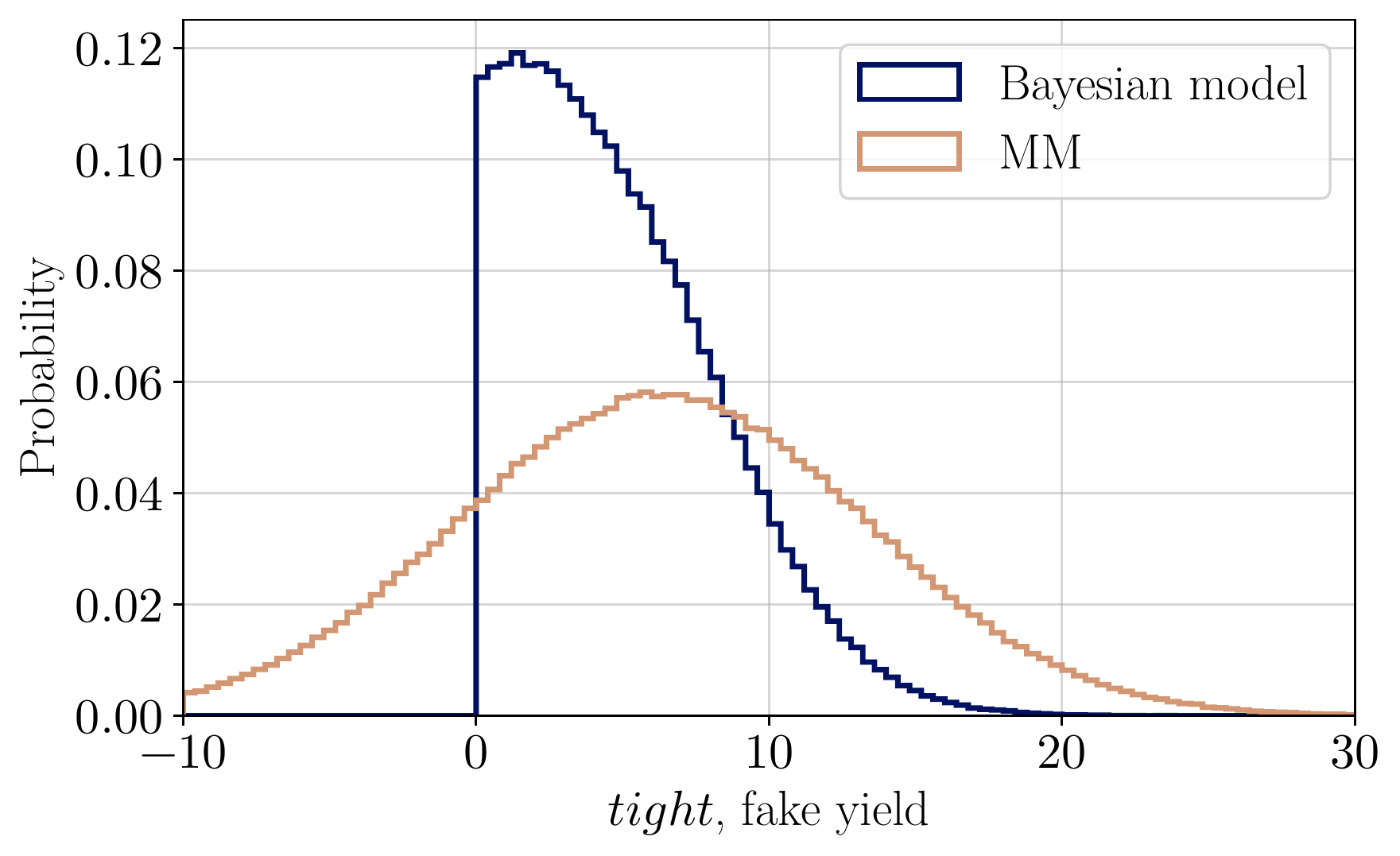}
        \subcaption{Case \Circled{5}.}
        \label{fig:case_5}
    \end{subfigure}
    \caption{Comparison of the original and improved matrix method to estimate fake-event yields in the \tight region. All distributions are sampled with $10^6$ samples. The Metropolis-Hastings algorithm is used for the Bayesian model posterior samples.}
    \label{fig:limit_case_comparisons}
\end{figure}

The results of the cases described above are presented as distributions in histograms in Figure~\ref{fig:limit_case_comparisons} and summarised in Table~\ref{tab:limit_case_table}. 
\begin{table}[H]
    \small
    \centering
    \caption{The quantitative results of the studies described above and illustrated in the histograms in Figure~\ref{fig:limit_case_comparisons}. The table shows the values of maximum probability (mode for the Bayesian approach and mean for the matrix method), as well as the median for the five limiting cases. The results for each of the cases are presented in columns. All uncertainties refer to the smallest \SI{68.27}{\percent} interval.}
    \label{tab:limit_case_table}
    \begin{tabular}{cc
        S[table-format=1.2(3)] 
        S[table-format=1.2(3)]
        S[table-format=1.2(3)]
        S[table-format=1.2(3)]
        S[table-format=1.2(3)]
      }
    \toprule
    \multicolumn{1}{c}{Model} & \multicolumn{1}{c}{Quantity} & {\Circled{1}} & {\Circled{2}} & {\Circled{3}} & {\Circled{4}} & {\Circled{5}} \\
    \midrule
    \multirow{2}{*}{Bayesian model} & mode & $\num{4.00}^{+3.50}_{-4.00}$ & $\num{0.00}^{+0.18}_{-0.00}$ & $\num{0.00}^{+0.12}_{-0.00}$ & $\num{0.08}^{+0.20}_{-0.08}$ & $\num{1.80}^{+4.60}_{-1.80}$ \\
    & median & $\num{5.25}$ & $\num{0.12}$ & $\num{0.79}$ & $\num{0.19}$ & $\num{4.41}$ \\ \midrule
    \multirow{2}{*}{matrix method} & mean & 10.00(2041) & 0.07(014) & -5.09(099) & 0.10(020) & 6.37(700) \\
    & median & $\num{10.00}$ & $\num{0.07}$ & $\num{-5.09}$ & $\num{0.10}$ & $\num{6.37}$ \\
    \bottomrule
    \end{tabular}
\end{table}
Each limit case is described in a column and the values of maximum probability of the distributions (mode for the Bayesian approach and mean for the matrix method) are given. The median is also given for both methods. Since the matrix method distribution follows a Gaussian distribution, mean value and median are identical. This is not the case for the Bayesian model, as the distributions presented in Figure~\ref{fig:limit_case_comparisons} are asymmetric. The uncertainties on the modal values refer to the upper and lower range of the smallest \SI{68.27}{\percent} intervals.

Each of the plots in Figure~\ref{fig:limit_case_comparisons} contains two histograms with the fake-event yield in the \tight regions. One histogram shows the distribution estimated with the original matrix method in orange, the distribution of the Bayesian model is shown in the other histogram in dark blue. The coloured regions indicating the smallest \SI{68.27}{\percent}, \SI{95.45}{\percent} and \SI{99.73}{\percent} intervals are omitted for clarity. The distributions computed with the original matrix method contain negative and thus unphysical contributions in all cases. 

The first example in Figure~\ref{fig:case_1} shows the expected unstable fake-event estimate with the matrix method as a broad distribution. The $x$-axis is truncated from \numrange{-10}{30} to also clearly illustrate the distribution of the Bayesian model. The estimate of the latter method is more stable, although the Bayesian approach leads to a broad plateau from $\ntf = 0$ up to $\ntf \approx 7$. 

In the second case, presented in Figure~\ref{fig:case_2}, the probability distribution of the matrix method decreases again towards zero. The Bayesian model does not show this feature and leads to a peak at zero but with a wider tail in the positive range.

Even in regions with limited validity of the original matrix method, the Bayesian model adequately describes the distribution of the fake leptons. An example for a region of limited validity is presented in Figure~\ref{fig:case_3}, where the distribution of the matrix method peaks in the negative range without any significant positive contribution. The Bayesian model shows a peak at zero with a steep slope, as expected.

In the fourth limit case in Figure~\ref{fig:case_4}, both methods produce narrow distributions with peaks and uncertainties of the same scale. Although the Bayesian model is evaluated at the prior bounds in this region, the estimate does not suffer. As expected, the distribution computed with the matrix method is very stable and accurate in terms of the mean value.

Since the efficiency uncertainties in the fifth case, given in Figure~\ref{fig:case_5}, are scaled with a factor of ten, the distributions of the two methods are wider, with a longer tail arising in the matrix method distribution. In this case, even with a positive estimated mean value of the matrix method, the values of maximum probability differ. 

In general, the Bayesian approach results in a narrower probability density for the fake-lepton yield, with zero probability for negative yields.

\FloatBarrier

\clearpage
\section{Conclusions}
\label{sec:conclusions}

The estimation of fake-lepton contributions is a key part of analyses that use prompt leptons. Since simulation-based estimations are impractical due to the small fake probabilities, data-driven techniques, such as the matrix method, are used instead.
Alternative approaches have been proposed~\cite{erich_varnes}, which make use of maximum likelihood estimates that provide reliable and stable fake-lepton estimates. In this paper, the likelihood approach was reformulated in Bayesian reasoning. Its origin has been developed to describe the collection and migration of \loose and \tight leptons, respectively, in a probabilistic fashion. The equality of the two approach has been shown. As an example application, the fake-lepton estimation of a top-quark measurement has been used to illustrate that the original matrix method results in non-negligible probabilities for negative event yields, while the improved method only predicts positive values. In five examples, the two methods have been compared, which show the strengths of the improved method over the original matrix method.
\appendix

\section{Appendix}

\subsection{Equality of the two approaches}\label{sec:appendix_proof}
\label{sec:equality}

In the following paragraph, it is proven that the two likelihoods from Eq.~\eqref{eqn:likelihood_first_principles} and Eq.~\eqref{eqn:lhmm_likelihood_2} are mathematically identical in the \loose frame of the likelihood matrix method.

Starting with the binomial theorem
\begin{equation}
  (x + y)^n = \sum_{k = 0}^{n} \binom{n}{k}x^{n-k}y^k\,,
  \label{eqn:binomial_theorem}
\end{equation}
the relation is applied to Eq.~\eqref{eqn:Nt_lhmm} and Eq.~\eqref{eqn:Nnt_lhmm}, resulting in
\begin{align}
  \begin{split}
    p(\Nt\,|\,\hat{\varepsilon}^{\mathrm{r}} \nu^{\mathrm{r}} + \hat{\varepsilon}^{\mathrm{f}} \nu^{\mathrm{f}}) &= \frac{(\hat{\varepsilon}^{\mathrm{r}} \nu^{\mathrm{r}} + \hat{\varepsilon}^{\mathrm{f}} \nu^{\mathrm{f}})^{\Nt}}{\Nt!} \text{e}^{-(\hat{\varepsilon}^{\mathrm{r}} \nu^{\mathrm{r}} + \hat{\varepsilon}^{\mathrm{f}} \nu^{\mathrm{f}})} \\
    &\stackrel{\eqref{eqn:binomial_theorem}}{=} \sum_{k = 0}^{\Nt}\binom{\Nt}{k} (\hat{\varepsilon}^{\mathrm{r}} \nu^{\mathrm{r}})^{\Nt - k} (\hat{\varepsilon}^{\mathrm{f}} \nu^{\mathrm{f}})^{k}\,\frac{\text{e}^{-(\hat{\varepsilon}^{\mathrm{r}} \nu^{\mathrm{r}} + \hat{\varepsilon}^{\mathrm{f}} \nu^{\mathrm{f}})}}{\Nt!}\,, \\
    p(N_{\text{nT}}\,|\,(1 - \hat{\varepsilon}^{\mathrm{r}})\nu^{\mathrm{r}} + (1 - \hat{\varepsilon}^{\mathrm{f}})\nu^{\mathrm{f}}) &= \frac{((1 - \hat{\varepsilon}^{\mathrm{r}})\nu^{\mathrm{r}} + (1 - \hat{\varepsilon}^{\mathrm{f}})\nu^{\mathrm{f}})^{N_{\text{L}} - \Nt}}{(N_{\text{L}} - \Nt)!}\text{e}^{-((1 - \hat{\varepsilon}^{\mathrm{r}})\nu^{\mathrm{r}} + (1 - \hat{\varepsilon}^{\mathrm{f}})\nu^{\mathrm{f}})} \\
    &\stackrel{\eqref{eqn:binomial_theorem}}{=} \sum_{n = 0}^{N_{\text{L}} - \Nt} \binom{N_{\text{L}} - \Nt}{n} ((1 - \hat{\varepsilon}^{\mathrm{r}})\nu^{\mathrm{r}})^{N_{\text{L}} - \Nt - n} \\
    &\hspace{1.9cm}\cdot((1 - \hat{\varepsilon}^{\mathrm{f}})\nu^{\mathrm{f}})^n\,\frac{\text{e}^{-((1 - \hat{\varepsilon}^{\mathrm{r}})\nu^{\mathrm{r}} + (1 - \hat{\varepsilon}^{\mathrm{f}})\nu^{\mathrm{f}})}}{(N_{\text{L}} - \Nt)!}\,,
  \end{split}
  \label{eqn:derivation_1}
\end{align}
with according parameterisations of \Nt and \Nnt. More details on the parameterisation are presented in Section~\ref{sec:Bayesian_model}.
Since both sums are independent of each other, the terms in Eq.~\eqref{eqn:derivation_1} can be multiplied to obtain Eq.~\eqref{eqn:lhmm_likelihood_2}
\begin{align}
  \begin{split}
    p(N_{\mathrm{T}}, N_{\mathrm{nT}}, \epsreal, \epsfake |\nu^{\mathrm{r}}, \nu^{\mathrm{f}}, \hat{\varepsilon}^{\mathrm{r}}, \hat{\varepsilon}^{\mathrm{f}}) &= \sum_{n = 0}^{N_{\text{L}} - N_{\text{T}}} \sum_{k = 0}^{N_{\text{T}}} \binom{N_{\text{T}}}{k}\binom{N_{\text{L}} - N_{\text{T}}}{n}(\hat{\varepsilon}^{\mathrm{r}} \nu^{\mathrm{r}})^{N_{\text{T}} - k} (\hat{\varepsilon}^{\mathrm{f}} \nu^{\mathrm{f}})^k \\
    &\hspace{1.2cm}\cdot\frac{((1 - \hat{\varepsilon}^{\mathrm{r}})\nu^{\mathrm{r}})^{N_{\text{L}} - N_{\text{T}} - n} ((1 - \hat{\varepsilon}^{\mathrm{f}})\nu^{\mathrm{f}})^n \mathrm{e}^{-\nu^{\mathrm{r}} - \nu^{\mathrm{f}}}}{N_{\text{T}}! (N_{\text{L}} - N_{\text{T}})!} \\
    &= \sum_{n = 0}^{N_{\text{L}} - N_{\text{T}}} \sum_{k = 0}^{N_{\text{T}}} (\hat{\varepsilon}^{\mathrm{r}})^{N_{\text{T}} - k}(\nu^{\mathrm{r}})^{N_{\text{L}} - n - k} (1 - \hat{\varepsilon}^{\mathrm{r}})^{N_{\text{L}} - N_{\text{T}} - n} \\
    &\hspace{1.2cm}\cdot\frac{(1 - \hat{\varepsilon}^{\mathrm{f}})^n (\hat{\varepsilon}^{\mathrm{f}})^k (\nu^{\mathrm{f}})^{n+k} \mathrm{e}^{-\nu^{\mathrm{r}} - \nu^{\mathrm{f}}}}{k!\,n!\,(N_{\text{T}} - k)! (N_{\text{L}} - N_{\text{T}} - n)!}\,.
  \end{split}
  \label{eqn:derivation_2}
\end{align}

Eq.~\eqref{eqn:derivation_2} can be compared to the likelihood resulting from the considerations in Eq.~\eqref{eqn:likelihood_first_principles}
\begin{align}
  \begin{split}
    p(N_{\mathrm{L}}, N_{\mathrm{T}}, \epsreal, \epsfake | \nu^{\mathrm{r}}, \nu^{\mathrm{f}}, \hat{\varepsilon}^{\mathrm{r}}, \hat{\varepsilon}^{\mathrm{f}}) &= \sum_{x = 0}^{N_{\text{L}}} \sum_{y = y_{\text{max}}}^{y_{\text{min}}} \binom{x}{y}\binom{N_{\text{L}} - x}{N_{\text{T}} - y} (\hat{\varepsilon}^{\mathrm{r}})^{N_{\text{T}} - y}(\nu^{\mathrm{r}})^{N_{\text{L}} - x} (\hat{\varepsilon}^{\mathrm{f}})^y (\nu^{\mathrm{f}})^x \\
    &\hspace{1.2cm} \cdot\frac{(1 - \hat{\varepsilon}^{\mathrm{r}})^{N_{\text{L}} - N_{\text{T}} - x + y} (1 - \hat{\varepsilon}^{\mathrm{f}})^{x - y} \mathrm{e}^{-\nu^{\mathrm{r}} - \nu^{\mathrm{f}}}}{x! (N_{\text{L}} - x)!} \\
    &= \sum_{x = 0}^{N_{\text{L}}} \sum_{y = y_{\text{max}}}^{y_{\text{min}}} (\hat{\varepsilon}^{\mathrm{r}})^{N_{\text{T}} - y}(\nu^{\mathrm{r}})^{N_{\text{L}} - x}(1 - \hat{\varepsilon}^{\mathrm{r}})^{N_{\text{L}} - N_{\text{T}} - x + y} \\
    &\hspace{1.2cm} \cdot\frac{(1 - \hat{\varepsilon}^{\mathrm{f}})^{x - y} (\hat{\varepsilon}^{\mathrm{f}})^y (\nu^{\mathrm{f}})^x \mathrm{e}^{-\nu^{\mathrm{r}} - \nu^{\mathrm{f}}}}{y!\,(x-y)!(N_{\text{T}} - y)! (N_{\text{L}} - N_{\text{T}} - x + y)!}\,.
  \end{split}
  \label{eqn:derivation_3}
\end{align}

The comparison of the coefficients and the corresponding exponents of Eq.~\eqref{eqn:derivation_2} and Eq.~\eqref{eqn:derivation_3} leads to
\begin{align}
  \begin{split}
    y &= k\, \\
    n &= x - y \Leftrightarrow x = n + k\,.
  \end{split}
  \label{statmod/eqn:x_and_y}
\end{align}
With these substitutions the lower sum index $x = 0$ in Eq.~\eqref{eqn:derivation_3} can be rewritten with the binomial coefficient $\binom{x}{y}$
\begin{equation}
    \binom{x}{y} \stackrel{\eqref{statmod/eqn:x_and_y}}{\to} \binom{n + k}{k}\,.
\end{equation}

Since the binomial coefficient is only defined for $n + k \geq k$~\cite{binomial_coefficient}, the lower limit for $n$ results in
\begin{equation}
  n \geq 0\,.
\end{equation}
The upper limit $y_{\mathrm{min}}$ is derived with the second binomial coefficient $\binom{N_{\mathrm{L}} - x}{N_{\mathrm{T}} - y}$ to
\begin{equation}
    \binom{N_{\mathrm{L}} - x}{N_{\mathrm{T}} - y} \stackrel{\eqref{statmod/eqn:x_and_y}}{\to} \binom{N_{\mathrm{L}} - n - k}{N_{\mathrm{T}} - k}\,.
\end{equation}

Using the same reasoning, this implies $N_{\mathrm{L}} - n - k \geq N_{\mathrm{T}} - k$, which leads to $n \leq N_{\mathrm{L}} - N_{\mathrm{T}}$.
The transformation of the sum over $y$ is done with the lower and upper limits of the first sum
\begin{align}
  \begin{split}
    y = k &= \mathrm{max}(0, n + k + N_{\mathrm{T}} - N_{\mathrm{L}}) \quad|\;n\;\text{cannot be larger than}\;N_{\mathrm{L}} - N_{\mathrm{T}} \\
    k &= \mathrm{max}(0, k) \\
    &\Rightarrow k \geq 0\,,
  \end{split}
\end{align}
while the upper limit follows as
\begin{align}
  \begin{split}
    k &= \mathrm{min}(n + k, N_{\mathrm{T}}) \quad|\;n\;\text{cannot be less than}\;0 \\
    k &= \mathrm{min}(k, N_{\mathrm{T}}) \\
    &\Rightarrow k \leq N_{\mathrm{T}}\,.
  \end{split}
\end{align}

Therefore, the ranges of $n$ and $k$ are defined by $0 \leq n \leq N_{\mathrm{L}} - N_{\mathrm{T}}$ and $0 \leq k \leq N_{\mathrm{T}}$, respectively, as given in Eq.~\eqref{eqn:derivation_2}.
\section*{Acknowledgements}

This work was supported by the Deutsche Forschungsgemeinschaft (DFG) through projects KR 4060/7-1 and KR 4060/13-1, the PUNCH4NFDI consortium supported by the DFG fund NFDI 39/1, the Studienstiftung des deutschen Volkes and the US Department of Energy (DoE).

\printbibliography
\end{document}